\documentclass[12pt,preprint]{article}
\usepackage{amssymb}
\usepackage{amsmath}
\usepackage{graphics}
\usepackage{epsfig}

\renewcommand{\vec}[1]{{\bf #1}}
\newcommand{\eqb}{\begin{equation}}
\newcommand{\eqe}{\end{equation}}
\newcommand{\dmb}{\begin{displaymath}}
\newcommand{\dme}{\end{displaymath}}

\newcommand{\eab}{\begin{eqnarray}}
\newcommand{\eae}{\end{eqnarray}}

\newcommand{\be}{\begin{equation}}
\newcommand{\ee}{\end{equation}}

\begin{document}

\begin{titlepage}
\begin{flushright}
KA-TP-16-2007 
\end{flushright}
\vspace{0.2cm}

\begin{center}
\Large{Black-Body Anomaly: Analysis of Temperature Offsets}
\vspace{1.0cm}

\large{Michal Szopa$^{**}$, Ralf Hofmann$^*$, Francesco Giacosa$^\dagger$, and 
\\Markus Schwarz$^*$}
\vspace{.5cm} 

{\small \begin{center}
{\em $\mbox{}^{**}$ Institut f\"ur Theoretische Physik\\ 
Universit\"at Heidelberg\\ 
Philosophenweg 16\\ 
69120 Heideberg, Germany}
\end{center}
\vspace{0.2cm} 

\begin{center}
{\em $\mbox{}^*$ Institut f\"ur Theoretische Physik\\ 
Universit\"at Karlsruhe\\ 
Kaiserstr. 12\\ 
76131 Karlsruhe, Germany}
\end{center}
\vspace{0.2cm}

\begin{center}
{\em $\mbox{}^\dagger$ Institut f\"ur Theoretische Physik\\ 
Universit\"at Frankfurt\\ 
Johann Wolfgang Goethe - Universit\"at\\ 
Max von Laue--Str. 1\\ 
60438 Frankfurt, Germany}
\end{center}}
\vspace{0.5cm}

\end{center}

\begin{abstract}

Based on the postulate that photon propagation is governed by a
dynamically broken SU(2) gauge symmetry (scale $\sim 10^{-4}\,$eV) we make 
predictions for temperature offsets due to a low-temperature 
(a few times the present CMB temperature) 
spectral anomaly at low frequencies. Temperature offsets are 
extracted from least-square fits of the anomalous black-body 
spectra to their conventional counterparts. We discuss statistical 
errors, compare our results with those obtained from calibration data of 
the FIRAS instrument, and point out that our 
predicted offsets are screened by experimental errors given the
frequency range used by FIRAS to perform their spectral fits. 
We also make contact with the WMAP observation by blueshifting their
frequency bands. Although our results
hint towards a strong dynamical component in the CMB dipole and an 
explanation of low-$l$ suppression it is important in view of its particle-physics implications
that the above postulate be verified/falsified by an 
independent low-temperature black-body precision experiment.

\end{abstract} 

\end{titlepage}

\section{Introduction}

The existence of the cosmic microwave background (CMB) was 
theoretically predicted a long time ago \cite{Gamow}. 
Its definite discovery, which gives tremendous support to the Hot Big Bang model, 
was announced almost twenty years later \cite{PenziasWilson}. An important 
result of the Cosmic Background Explorer (COBE) mission and 
in particular its Far-Infrared Absolute Spectrometer (FIRAS) was the 
observation of a black-body 
spectral shape of the CMB radiation in the frequency 
range from 1 or 2\,cm$^{-1}$ to 20\,cm$^{-1}$ with temperature uncertainties of at most a 
few milli Kelvin \cite{FIRASdoc,FIRAS1994I,FIRAS1994II,FIRAS1999}. 

A successor satellite experiment, the Wilkinson Microwave 
Anisotropy Probe (WMAP) \cite{WMAPdoc,Hinshaw2006}, recorded the full-sky maps of temperature 
variations by observing this near-to-perfect 
black-body spectrum within five fixed frequency bands, $\nu=23, 33, 41, 61,$ and 94\,GHz, picked 
in such a way as to minimize the 
contamination of the CMB signal by the galactic and the 
solar-system foregrounds. 

Based on the postulate that photon {\sl propagation} is described by 
an SU(2) Yang-Mills theory \cite{Hofmann2005,Hofmann2005BP} a spectral black-body (BB) anomaly was predicted to 
occur in the low-temperature, low-frequency realm
\cite{SHG2006,SHG2006BB}. Namely, for frequencies below a
temperature-dependent critical value the BB spectral intensity vanishes
because of photon screening. Above this
critical frequency there is a small range of antiscreening. Both effects
are due to the nonabelian nature of the theory, see \cite{SHG2006BB} and
Sec.\,\ref{esu2}. Because of its association with 
the CMB, whose present temperature determines the Yang-Mills scale to 
$\Lambda\sim 10^{-4}\,$eV$\sim T_{\tiny\mbox{CMB}}=2.73$\,K, the name
SU(2)$_{\tiny\mbox{CMB}}$ was introduced.                            

The thermodynamics of an SU(2) (or SU(3)) Yang-Mills theory is
approached by addressing the (microscopically highly complex) dynamics 
of the ground state in terms of an effective theory obtained by
coarse-graining 
the topologically nontrivial sector 
\cite{RH2005B}. SU(2) Yang-Mills thermodynamics comes in three phases: 
A deconfining one for temperatures $T\ge T_c\sim\Lambda$, a preconfining one
for $T\sim\Lambda$, and a confining one for $T<\Lambda$, for more detail
see Sec.\,\ref{esu2}. 

As for the modification of the 
radiation law of a black body by nonabelian effects the 
deconfining phase of SU(2)$_{\tiny\mbox{CMB}}$ is 
relevant. For an analysis of the viability of SU(2)$_{\tiny\mbox{CMB}}$
see \cite{GH2005}. 
Other interesting predictions, such as the duration of the present cosmological epoch of supercooling 
\cite{GH2005}, require quantitative knowledge about the 
preconfining phase. In that phase the photon nonperturbatively 
acquires a (Meissner) mass. 

In connection with the onset of a superconducting ground state 
let us consider a photon gas contained in a small spatial region immersed in the CMB, a satellite for example. When cooling this gas below 
$T_{\tiny{\mbox{CMB}}}$ the associated ground state would be preconfining 
(massive photon) if it were not for the 
large-range correlation in the surrounding CMB 
ground state which prevents this transition from occurring 
(supercooled situation). The physical effect associated 
with tunneling transitions between the two ground states  
is an increase of the effective number of 
photon polarizations. The effect depends on the ratio 
between correlation length and 
linear dimension of the cold-spot volume. For details see
Sec.\,\ref{Firas} below. 

The occurrence of a BB anomaly for $T>T_{\tiny\mbox{CMB}}$ 
and low frequencies has a profound impact on 
the CMB temperature-temperature correlations at large angles. 
The suppression of the latter after subtracting the 
CMB monopole and dipole and the violation of statistical isotropy at 
low $l$ \cite{Schwarz2007} is likely to be understood by 
this anomaly. In \cite{SH2007} a dynamically generated component to 
the CMB dipole was considered to be due to the BB anomaly.

The purpose of the present paper is an analysis of the present 
experi\-men\-tal/ob\-ser\-vational situation concerning the afore-mentioned BB
anomaly. We concentrate on two satellite-based 
experiments. First, we explore the consequences of the BB 
anomaly for offsets between the known wall temperature and 
the fitted temperature of radiation when assuming the latter to be spectrally distributed 
according to Planck's law. This was the strategy used in the temperature
calibration stage 
of FIRAS. The calibration was performed both on Earth and in orbit with differing results. Here we would like to compare our theoretical 
results with their in-flight data. Apart from an interesting signature at 
$T<T_{\tiny\mbox{CMB}}$ and a large negative (global) offset for
$T>T_{\tiny\mbox{CMB}}$, which we discuss in view of
SU(2)$_{\tiny\mbox{CMB}}$, we conclude that for fits within the above-quoted FIRAS frequency 
range the effect of the theoretically predicted BB anomaly is too small to rise above the 
experimental errors. 

Second, we would 
like to make contact with the WMAP satellite (temperature fits within
the above-quoted but blueshifted frequency bands). Since blueshifted spectra are 
related to the angular resolution in the temperature-temperature correlation 
function we are in a position to semi-quantitatively compare
observationally determined temperature offset for $T>T_{\tiny\mbox{CMB}}$ with theory.  

The article is organized as follows. For the benefit of the reader, we quote 
a number of essential results for deconfining SU(2) Yang-Mills thermodynamics 
in Sec.\,\ref{esu2}. In Sec.\,\ref{OSTP} we first describe the 
relevant results for temperature offsets in the low-temperature realm 
as measured and observed by FIRAS and WMAP, respectively. 
Within the setting of FIRAS we then perform 
spectral fits to theoretical predictions based on
SU(2)$_{\tiny\mbox{CMB}}$, and we discuss our results. As a next step, we perform an analogous program 
for the blueshifted WMAP frequency bands and discuss these results in
view of the low TT multipoles in the CMB. Finally, we extract temperature offsets
from integrated spectra and point out the difference compared 
with spectral fits. In Sec.\,\ref{SC} we summarize our results and conclude.     

\section{Effective SU(2) Yang-Mills thermodynamics\label{esu2}}

In this section we briefly describe the effective theory for
SU(2) Yang-Mills thermodynamics. The basic ingredient is an adjoint
scalar field $\phi $, 
which represents (part of) the thermal ground state emerging from a
spatial average over interacting calorons and anticalorons of topological
charge modulus $|Q|=1$. The scalar field $\phi$ is an inert background
field to the dynamics of topologically trivial, coarse-grained gauge fields $%
a_{\mu }$ entering in the effective field strength $G_{\mu \nu
}^{a}=\partial _{\mu }(a_{\nu }^{a})-\partial _{\nu }(a_{\mu
}^{a})-e\varepsilon ^{abc}a_{\mu }^{b}a_{\nu }^{c}$. The temperature
dependent effective coupling $e=e(T)$ follows from thermodynamical
selfconsistency, see below. The effective Lagrangian for the description of
SU(2)-Yang-Mills thermodynamics in the deconfining phase ($T>T_{c}=\lambda
_{c}\Lambda /2\pi $, $\lambda _{c}=13.87,$ $\Lambda$ the Yang-Mills
scale) and in unitary gauge reads \cite{RH2005B}: 
\begin{equation}
\mathcal{L}_{{\tiny \mbox{dec-eff}}}^{u.g.}=\frac{1}{4}\left( G_{E}^{a,\mu
\nu }[a_{\mu }]\right) ^{2}+2e(T)^{2}\left\vert \phi \right\vert ^{2}\left(
\left( a_{\mu }^{(1)}\right) ^{2}+\left( a_{\mu }^{(2)}\right) ^{2}\right) +%
\frac{2\Lambda _{E}^{6}}{\left\vert \phi \right\vert ^{2}}\,.
\label{lageffdec}
\end{equation}%
The modulus of the adjoint scalar field $\left\vert \phi \right\vert $
depends on the Yang-Mills scale $\Lambda$ and on temperature $T$ as $%
\left\vert \phi \right\vert =\sqrt{\frac{\Lambda ^{3}}{2\pi T}}$. The
quantity $\left\vert \phi \right\vert ^{-1}$ is the minimal length
(inherent resolution) down to which the thermalized 
system appears spatially homogeneous. In other words,
the spatial average over interacting calorons and anticalorons
selfconsistently saturates below this length scale. The quantum fluctuations $%
a_{\mu }^{(1,2)}$ are massive  in a temperature dependent way, $%
m^{2}=m(T)^{2}=4e^{2}\left\vert \phi \right\vert ^{2}$, while the gauge mode 
$a_{\mu }^{(3)},$ stays massless. In \cite{Hofmann2005} the postulate
was made to identify this massless mode with the photon. Notice
that the dynamical gauge symmetry breaking $SU(2)\rightarrow
U(1) $ occurs by virtue of nontrivial topology: $\phi$ is an adjoint Higgs field.

We work with the following dimensionless quantities%
\begin{equation}
\overline{\rho }=\frac{\rho }{T^{4}}\,,\ \text{ }\overline{p}=\frac{p}{T^{4}}%
\,,\ \text{ }\lambda =\frac{2\pi T}{\Lambda }\,,\ \text{ }a(\lambda )=\frac{%
m(T)}{T}=2\frac{e(T)}{T}\left\vert \phi \right\vert 
\end{equation}%
where $\rho $ and $p$ are the energy density and the pressure associated
with the partition function of the Lagrangian (\ref{lageffdec}), and the function $a=a(\lambda )$ is introduced
for later use. On the one-loop level, the energy density and pressure $\rho$ and $p$ are a sum of
three contributions 
\begin{equation}
\rho =\rho _{3}+\rho _{1,2}+\rho _{gs}\,,\ \ \text{ }p=p_{3}+p_{1,2}+p_{gs}%
\,,  \label{rhop}
\end{equation}%
where the subscript 1,2 is understood as a sum over the two massive modes $%
a_{\mu }^{(1,2)}$, the subscript 3 refers to the massless mode $a_{\mu
}^{(3)}$, and the subscript `$gs$' labels the ground-state contribution.
When expressing $\overline{\rho }$ and $\overline{p}$ as functions of the
dimensionless temperature $\lambda $, one obtains at one-loop (accurate on
the 0.1\%-level \cite{Hofmann2005,SHG2006}):%
\begin{equation}
\overline{\rho }_{3}=2\frac{\pi ^{2}}{30},\text{ }\overline{\rho }_{1,2}=%
\frac{3}{\pi ^{2}}\int_{0}^{\infty }dx\frac{x^{2}\sqrt{x^{2}+a^{2}}}{e^{%
\sqrt{x^{2}+a^{2}}}-1},\text{ }\overline{\rho }_{gs}=\frac{2(2\pi )^{4}}{%
\lambda ^{3}}.  \label{rhoad}
\end{equation}%
\begin{equation}
\overline{p}_{3}=2\frac{\pi ^{2}}{90},\text{ }\overline{p}_{1,2}=-\frac{3}{%
\pi ^{2}}\int_{0}^{\infty }x^{2}dx\ln \left( 1-e^{-\sqrt{x^{2}+a^{2}}%
}\right) ,\text{ }\overline{p}_{gs}=-\text{ }\overline{\rho }_{gs}\,.
\label{pad}
\end{equation}%
The ground-state contribution acts like a temperature 
dependent bag constant with negative pressure. Imposing the validity of the thermodynamical Legendre
transformation 
\begin{equation}
\rho =T\frac{dP}{dT}-P\iff \overline{\rho }=\lambda \frac{d\overline{p}}{%
d\lambda }+3\overline{p}  \label{tsc}
\end{equation}%
and substituting the expressions (\ref{rhoad})-(\ref{pad}) into (\ref{tsc}),
we arrive at the following differential equation for $a=a(\lambda )$:%
\begin{eqnarray}
1 &=&-\frac{6\lambda ^{3}}{(2\pi )^{6}}\left( \lambda \frac{da}{d\lambda }%
+a\right) aD(a),\text{ }  \label{aeq} \\
D(a) &=&\int_{0}^{\infty }dx\frac{x^{2}}{\sqrt{x^{2}+a^{2}}}\frac{1}{e^{%
\sqrt{x^{2}+a^{2}}}-1},\text{ }a(\lambda _{in})=0\,.
\end{eqnarray}%
For a sufficiently large initial value $\lambda_{in}$ the solution for $%
a(\lambda )$ is independent on $\lambda _{in}$: a low-temperature attractor
with a logarithmic pole at $\lambda _{c}=13.87$ is seen to exist. The
effective coupling is given as $e=e(\lambda )=a(\lambda )\lambda ^{3/2}/4\pi 
$, and $e$ exhibits a plateau $e=\sqrt{8}\pi $ for $\lambda \gg \lambda _{c}$.
In fact, $a(\lambda )=\frac{8\sqrt{2}\pi ^{2}}{\lambda ^{3/2}}$ is a
solution of the differential equation (\ref{aeq}) for $a\ll 1$, that is, for 
$\lambda \gg \lambda _{c}.$ For plots and the discussion of the
thermodynamical quantities we refer to \cite{Hofmann2005}. The effective
theory also leads to a linear growth in $T$ of $\left\langle
\theta _{\mu \mu }\right\rangle _{T}=\varepsilon -3p$ for large temperatures 
\cite{lg} in agreement with lattice simulations
\cite{lattice}. 

Depending on its momentum, the massless excitation (photon) suffers screening 
or antiscreening due to radiative corrections in the deconfining
phase. This follows from a calculation of the one-loop polarization
tensor for this mode \cite{SHG2006}. While these effects are negligible for
$\lambda\gg \lambda_c$ there are sizable modifications of the dispersion
law for $\lambda_c\lesssim\lambda$ and for $|\vec{p}|/T\sim 0.1$ where
$\vec{p}$ is the massless mode's spatial momentum. This, in turn,
implies the occurrence of the low-temperature, low-frequency BB anomaly \cite{SHG2006BB}.

The logarithmic pole of $e(\lambda )$ at $\lambda _{c}=$ $13.87$ implies
diverging masses, thus a thermodynamical decoupling, and the stalling of
radiative corrections. At the same time
screened monopoles, which were liberated by the dissociation of
large-holonomy calorons, start to form a condensate. For $2\pi \lambda _{c,M}/\Lambda =T_{c,M}\leq T\leq T_{c}=2\pi
\lambda _{c}/\Lambda ,$ $\lambda _{c,M}=11.57$ the system is in the
so-called preconfining (magnetic) phase: the remaining gauge mode $a_{\mu
}^{(3)}$ acquires mass by interacting with the monopole
condensate. Since there is a gap in the energy density at $\lambda _{c}$
the system will remain in a supercooled situation even for temperatures
below $\lambda_{c}$. Apart from the small effect of tunneling between
the deconfining and the preconfining ground state 
the supercooled (yet deconfining) ground state completely decouples from
its massless excitations. Based on
this theoretical input the possibility that the photon becomes Meissner
massive during a future cosmological epoch has been investigated in
\cite{GH2005}. For a temperature smaller than $\lambda _{c,M}$ 
the system is in a fully confining phase: Then the excitations
(single or selfintersecting center-vortex loops) are massless or massive
spin-1/2 fermions.

\section{Observational situation and 
theoretical predictions\label{OSTP}}

\subsection{Brief summary of present experimental situation}

\subsubsection{Temperature calibration data: FIRAS } 

The FIRAS instrument was part of COBE's 
set of instruments to measure to what extent the CMB spectrum possesses a BB shape \cite{FIRASdoc,FIRAS1994I,FIRAS1994II,FIRAS1999}. 

The part of the mission, which is of interest to the present work, is the calibration 
stage since SU(2)$_{\tiny\mbox{CMB}}$ does not predict any deviation
from a perfect, conventional BB spectrum at $T=T_{\tiny\mbox{CMB}}$. The
experimental procedure of calibration surely is highly complex. In spite
of this we believe that a number of interesting features can be seen. 

\begin{figure}[tbp]
\begin{center}
\leavevmode
\leavevmode
\vspace{10.5cm} \includegraphics{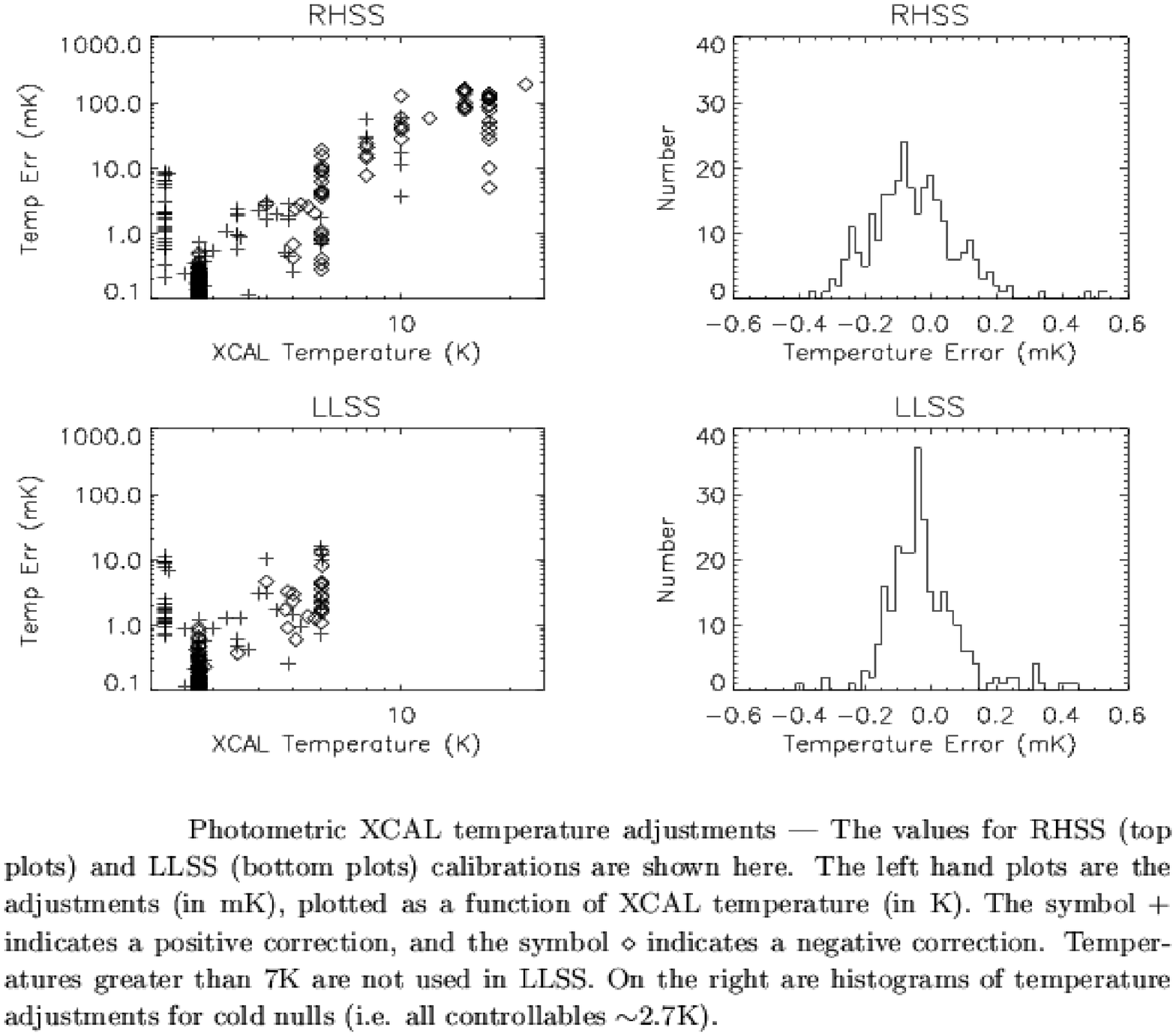}
\end{center}
\caption{Temperature offset as measured during the FIRAS in-flight calibration, figure taken from \protect{\cite{FIRASdoc}}.}
\label{Fig-0}
\end{figure}

Let us now sketch the situation. A
close-to-ideal\footnote{Emissivity close to unity.} 
calibrator BB (XCAL) was placed in front of the microwave antenna (horn) 
to simulate an ideal BB spectrum of an externally adjustable temperature $T_{\tiny\mbox{XCAL}}$. To calibrate the bolometers 
the spectral power of radiation as emitted from XCAL was
measured. Subsequently, a temperature $T_{\tiny\mbox{rad}}$ was 
extracted by fitting this data to a Planck curve. This was done both on
Earth and in orbit. 
The terrestial calibration yielded a much lower offset between 
$T_{\tiny\mbox{XCAL}}$ and $T_{\tiny\mbox{rad}}$ than the in-flight calibration did \cite{FIRAS1994I}. Defining the offset by 
$\delta T\equiv T_{\tiny\mbox{rad}}-T_{\tiny\mbox{XCAL}}$, $\delta T$
was minus a few mK (see pages pp. 45--47 
in \cite{FIRASdoc}), $T_{\tiny\mbox{XCAL}}$ ranging as $2.7\,\mbox{K}\le T_{\tiny\mbox{XCAL}}\le 6\,$K. Interestingly, a {\sl positive} 
$\delta T$ of order mK was extracted at $T_{\tiny\mbox{XCAL}}=2.2\,$K, see Fig.\,\ref{Fig-0} (taken from p.\,45 of Ref.\,\cite{FIRASdoc}).
\subsubsection{Results for CMB temperature-temperature correlation: WMAP}

The WMAP satellite was (among others) designed to 
observe the angular spectrum of the monopole and the dipole
\footnote{The dipole has an amplitude of $\left.\delta T\right|_{l=1}=3.358\pm 0.017\,$mK \cite{Hinshaw2006}.} 
subtracted TT correlator of CMB temperature fluctuations with 
unprecedented precision. The temperature map underlying this is obtained by comparing the intensity measured for 
a given direction on the sky and for a given frequency with the respective 
intensity of an ideal BB. Frequencies considered are contained in one of the bands centered at $\nu=23, 33,
41, 61,$ and 94\,GHz. As explained in detail in \cite{Schwarz2007}, temperature-temperature 
correlations extracted from the CMB map practically vanish at large angles ($\theta>60$ degrees) when compared 
to a prediction based on a scale-invariant spectrum of primordial density perturbations. Also, the low multipoles seem to 
be statistically correlated which casts doubts on the validity 
of the usual assumption of statistical isotropy applied when 
extracting the power of these multipoles.

\subsection{Comparison between SU(2)$_{\tiny\mbox{CMB}}$ and U(1)} 

Here we would like to present results for the BB spectra and temperature 
offsets $\delta T$ as they are predicted when replacing the conventional
gauge group for photon 
{\sl propagation} U(1) by SU(2)$_{\tiny\mbox{CMB}}$. 
We define $\delta T\equiv \bar{T}-T$ where the temperature $\bar{T}$ is extracted 
from a fit of data to the conventional 
Planck spectrum. This data represents a modified Planck spectrum as generated 
by SU(2)$_{\tiny\mbox{CMB}}$ at temperature $T$ 
($T$ plays the role of $T_{\tiny\mbox{XCAL}}$ at FIRAS). Both $\bar{T}$
and $T$ are measured in K. The theoretical basis for the 
postulate SU(2)$_{\tiny\mbox{CMB}}\stackrel{\tiny\mbox{today}}=$U(1)$_Y$ 
and some consequences thereof are discussed in Sec.\,\ref{esu2}, see
also \cite{Hofmann2005BP,SHG2006,SHG2006BB,RH2005B,GH2005,SH2007}.

\subsubsection{Embedding of a black body}  

Before discussing the results of the FIRAS calibration and the WMAP 
observation we would like to point out the conditions under 
which the BB anomaly can be detected experimentally. 

Strictly speaking, the occurrence of the BB anomaly presumes the absence 
of conventional, electrically charged matter 
within the BB cavity (SU(2)$_{\tiny\mbox{CMB}}$ describes photon
propagation only.) A good approximation to this situation is difficult to 
achieve in terrestial experiments. It is, however, helpful that the
effect is predicted to occur for low temperatures where the excitability of 
residual-gas molecules in the cavity by the BB radiation is very limited. For a measurement 
in space there is much better vacuum which suggest a cleaner spectral
signature. 

Let us now address the issue of how temperature offsets $\delta T$ are
affected by the embedding of the BB into its 
surroundings. First we consider the situation of FIRAS. Here a
BB is immersed into the surrounding CMB radiation. Compared to the
calibration on Earth a larger negative offset of 
about 4\,mK of the radiation temperature as compared to the wall 
temperature of XCAL was seen during the in-flight calibration
($T_{\tiny\mbox{XCAL}}>T_{\tiny\mbox{CMB}}$).  

In our opinion this effect is due to the finite volume of order 1\,m$^3$ of the FIRAS instrument and the fact
that the correlation length $l(T)$ in the thermal ground state of
SU(2)$_{\tiny\mbox{CMB}}$ is much
larger\footnote{The value $l(T_{\tiny\mbox{CMB}})\sim 1\,\mbox{km}$ 
assumes a value $e\sim 10^5$ for the effective gauge coupling at the
onset of monopole condensation, for a discussion see \cite{GH2005}.} than the linear dimension of the BB cavity at
$T_{\tiny\mbox{CMB}}$: $l(T_{\tiny\mbox{CMB}})\sim 1\,\mbox{km}$. In this 
case, the ground state within the satellite is perceptibly
influenced by the ground state of the surroundings because of the large
$l$. Going from outside to inside across the BB wall, $l$ is a steeply 
falling function of distance since the mass of a screened monopole
varies rapidly with temperature in the vicinity of
$T_{\tiny\mbox{CMB}}$. Radiation emitted from the walls of the 
BB partially acknowledges the presence of this transition 
region by loosing part of its energy to the ground 
state outside the BB. This, in turn, implies a slight reduction of the 
radiation temperature inside the BB cavity. 

\begin{figure}[tbp]
\begin{center}
\leavevmode
\leavevmode
\vspace{6.7cm} \includegraphics{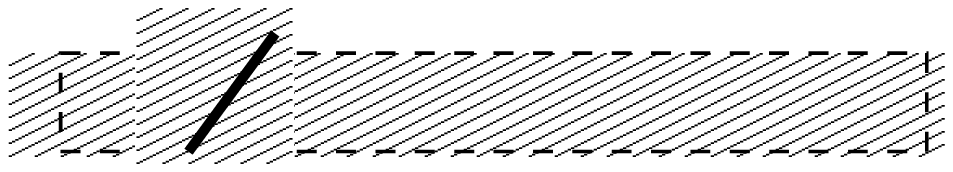}
\end{center}
\caption{Dependence of the offset $\delta T$ on the effective number of polarizations 
$F$ at $T=2.2\,$K. The shaded band signals the approximate region for
$\delta T$ as measured by FIRAS.}
\label{Fig-4}
\end{figure}

But what about the {\sl positive} temperature offset $\delta T$ of a few mK 
at $T=2.2\,$K, see Fig.\,\ref{Fig-0}? We like to 
interprete this effect as an increase of the {\sl effective} number of photon polarizations 
$F$. Namely, in the case of an infinite BB volume and at $T=2.2\,$K SU(2)$_{\tiny\mbox{CMB}}$ 
would be in its confining phase. In the intermediate, preconfining phase 
the photon would be massive with three polarizations. Since the FIRAS BB
has a finite volume and since the ground state of the surrounding Universe has practically an 
{\sl infinite} correlation length ($l_c\sim 1\,\mbox{km}$) we expect the
ground state of the BB cavity to be {\sl almost} identical to that of
the Universe as a whole. 
That is, the ground state of the BB cavity {\sl almost} decouples from 
the excitations, see Sec.\,\ref{esu2}. A deviation from  
this situation by tunneling between the pre- and deconfining ground 
state, which would mean an onset of superconductivity
(monopoles are electrically charge w.r.t U(1)$_Y$) and thus the excitation
of a longitudinal photon polarization, 
is described by a function $f$ of the cavity volume and the correlation 
length $l$. A quantitative grasp of $f$ would need information about the 
deviation from the infinite-volume limit considered in
\cite{Hofmann2005}. This, however, is beyond the scope of the present
work. Due to tunneling a nonvanishing but small probability exists for a photon to
possess three instead of two polarizations. As far as BB radiation is concerned, we model this situation by 
introducing an average number $F$ of photon polarizations which multiplies the conventional 
Planck spectrum by $F/2$. In Fig.\,\ref{Fig-4} a plot of the (positive) 
$\delta T$ as a function of $F$ at $T=2.2\,$K is depicted. According to Fig.\,\ref{Fig-4} a typical offset of about 5\,mK, 
compare with Fig.\,\ref{Fig-0}, corresponds to about a 1\%-increase of
the effective number of photon polarizations.   

\begin{figure}[tbp]
\begin{center}
\leavevmode
\leavevmode
\vspace{6.5cm} \includegraphics{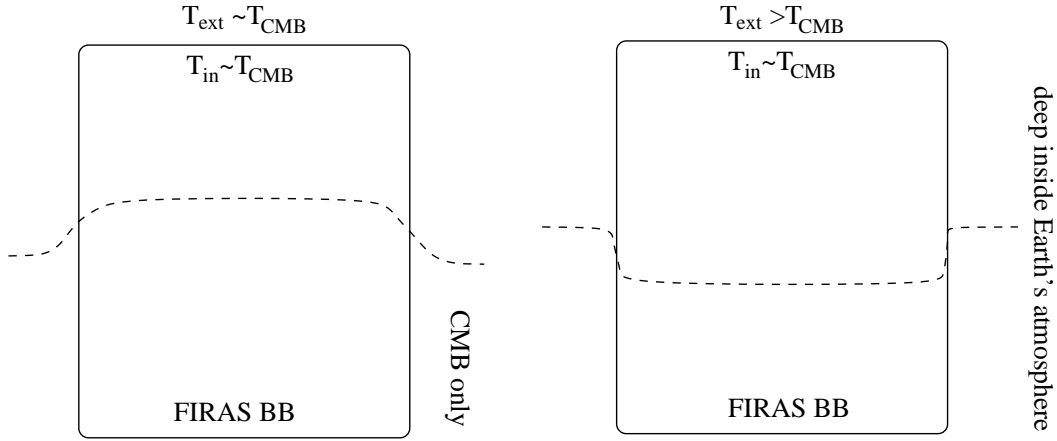}
\end{center}
\caption{Sketch of the influence of the surrounding's ground state on
  the ground state inside the BB cavity. A large correlation length
  corresponds to in-flight conditions (left panel; small gradient 
in ground-state energy density represented by dashed line) and a small correlation
  length to a lab placed on Earth (right panel; large gradient).} 
\label{Fig-3-C}
\end{figure}

Let us now discuss the situation of a terrestial low-temperature BB
experiment. Here the surroundings are at a much higher
temperature than $T_{\tiny\mbox{CMB}}$ implying a correlation length $l$
of the external ground state of $l<1\,\mbox{cm}$ \cite{RH2005B}. This
situation is sketched in the right panel of Fig.\,\ref{Fig-3-C}. Hence in the
case of terrestrial FIRAS calibration the warmer ground state of the
surroundings had practically no effects on 
the BB radiation temperature. This would qualitatively explain why the
terrestial calibration has seen a much smaller temperature offset
$\delta T$ than the in-flight calibration did.  

\subsubsection{Experiment and SU(2)$_{\tiny\mbox{CMB}}$: FIRAS\label{Firas}}  

\begin{figure}[tbp]
\begin{center}
\leavevmode
\leavevmode
\vspace{4.5cm} \includegraphics{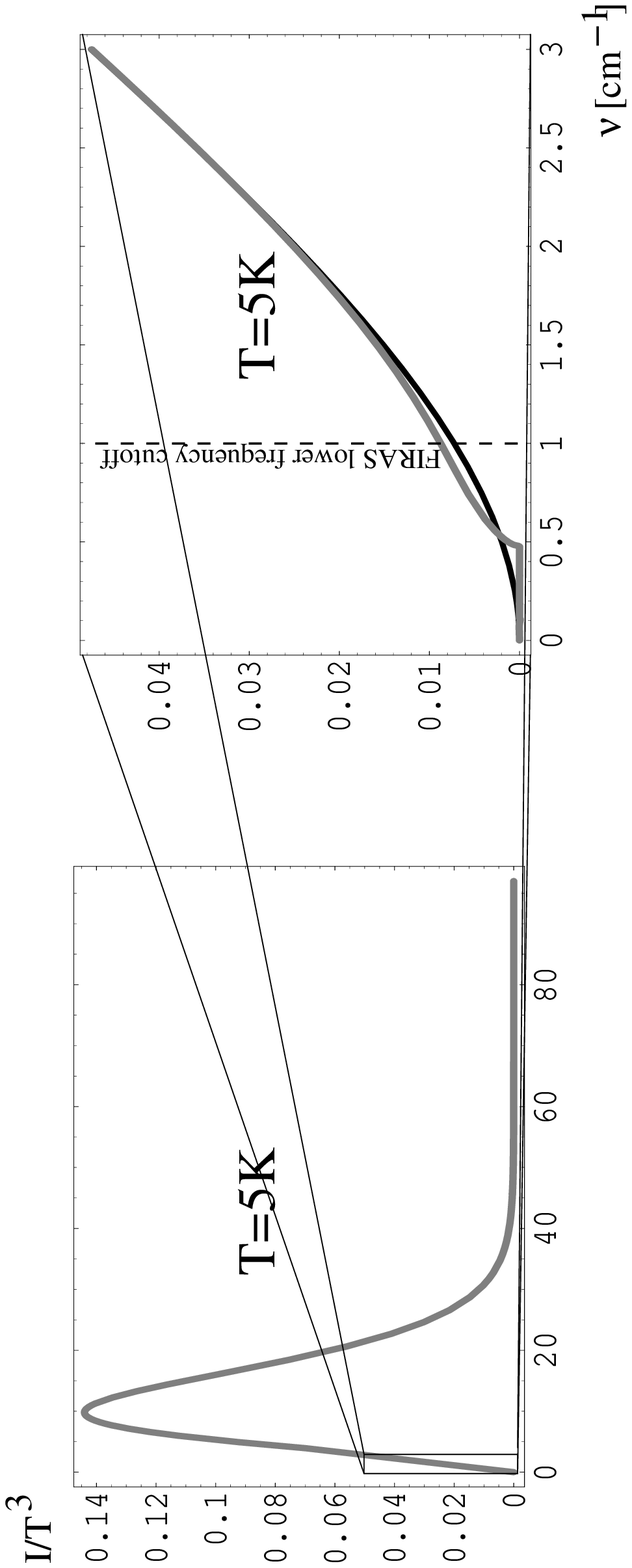}
\end{center}
\caption{Dimensionless spectral intensity of black-body 
radiation at $T=5\,$K as described by SU(2)$_{\tiny\mbox{CMB}}$ (gray curve) and 
standard U(1) (black curve). The unit of frequency $\nu$ 
is cm$^{-1}$. The left (right) panel depicts the spectrum 
for $0\le\nu\le 97$\,cm$^{-1}$ (for $0\le\nu\le 3$\,cm$^{-1}$).}
\label{Fig-1}
\end{figure}

The unit of frequency $\nu$ used in the FIRAS experiment is $1\,\mbox{cm}^{-1}$.  
Since FIRAS has performed most of its BB fits in the range
$\nu_{\tiny\mbox{low}}\equiv 1\,
\mbox{cm}^{-1}\le\nu\le 20\,\mbox{cm}^{-1}\equiv \nu_{\tiny\mbox{up}}$ we 
will also perform our fits within this range for temperatures up 
to $T=4\,$K. The reason why we do not consider a fixed value
$\nu_{\tiny\mbox{up}}=20\,\mbox{cm}^{-1}$ 
for temperatures $T$ higher than 4\,K 
is the following. For $T>4\,$K the fits do not 
saturate when varying $\nu_{\tiny\mbox{up}}$ in the vicinity of $\nu_{\tiny\mbox{up}}=20\,\mbox{cm}^{-1}$. 
Thus we prescribe $\nu_{\tiny\mbox{up}}=\frac{10}{\mbox{K$\cdot$cm}}\,T$ which is deep inside 
the saturation regime\footnote{We have observed that the fits saturate well already 
for $\nu_{\tiny\mbox{up}}$ being the maximum of the BB spectrum. So 
the value $\nu_{\tiny\mbox{up}}=\frac{10}{\mbox{K$\cdot$cm}}\,T$, which
is far to the right of the maximum, is very safe.}. 
In Fig.\,\ref{Fig-1} plots of 
the dimensionless (in natural units with $c=\hbar=k_B=1$) BB spectra are given for 
both the conventional case and the case of SU(2)$_{\tiny\mbox{CMB}}$. Notice the regime 
in frequency of excess (suppression) of spectral power to the right
(left) of the critical frequency $\nu^{*}\sim 0.5\,$cm$^{-1}$. 
Suppression is due to photon screening while the excess is caused by antiscreening \cite{SHG2006,SHG2006BB}. 
Notice also that FIRAS could only observe antiscreening since its value for $\nu_{\tiny\mbox{low}}=1\,\mbox{cm}^{-1}$ is to the right of $\nu^{*}$. 
A Planck-curve fit to the modified spectrum leads to a negative value of
$\delta T$. This is because by 
Wien's displacement law a temperature $\bar T<T$ is required to account
for the larger spectral power at low frequencies. 
The situation that $\nu^{*}<\nu_{\tiny\mbox{low}}=1\,\mbox{cm}^{-1}$ 
changes for higher temperatures, see Fig.\,\ref{Fig-2}. For example, at
$T=20\,$K we have $\nu^{*}>\nu_{\tiny\mbox{low}}$. However, as we will 
see below, the offset $\delta T$ remains negative for $T\ge 20\,$K. This is explained by the fact that the spectral
weight of the screening regime ($\nu<\nu^{*}$) is much lower than that
belonging to the regime of antiscreening.
\begin{figure}[tbp]
\begin{center}
\leavevmode
\leavevmode
\vspace{4.5cm} \includegraphics{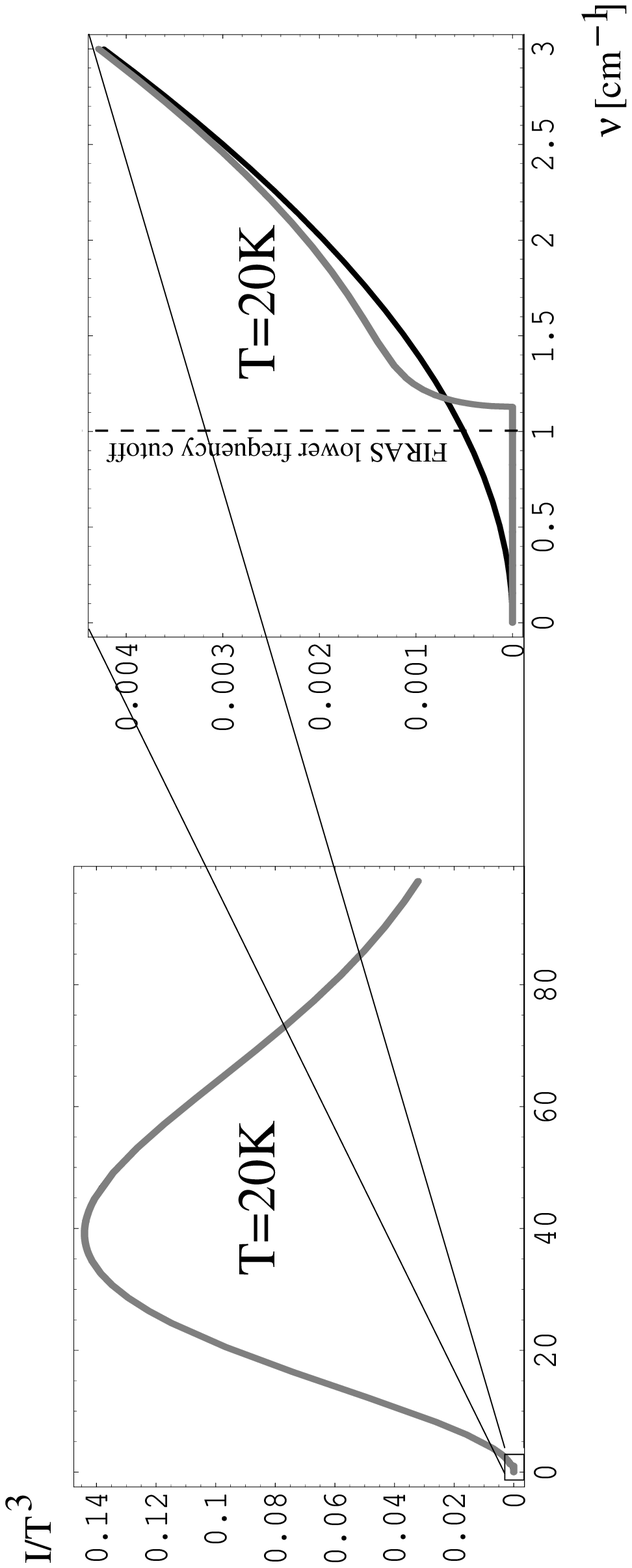}
\end{center}
\caption{Dimensionless spectral intensity of black-body 
radiation at $T=20\,$K as described by SU(2)$_{\tiny\mbox{CMB}}$ (gray curve) and 
standard U(1) (black curve). The unit of frequency $\nu$ 
is cm$^{-1}$. The left (right) panel depicts the spectrum 
for $0\le\nu\le 97$\,cm$^{-1}$ (for $0\le\nu\le 3$\,cm$^{-1}$).}
\label{Fig-2}
\end{figure}

\begin{figure}[tbp]
\begin{center}
\leavevmode
\leavevmode
\vspace{4.5cm} \includegraphics{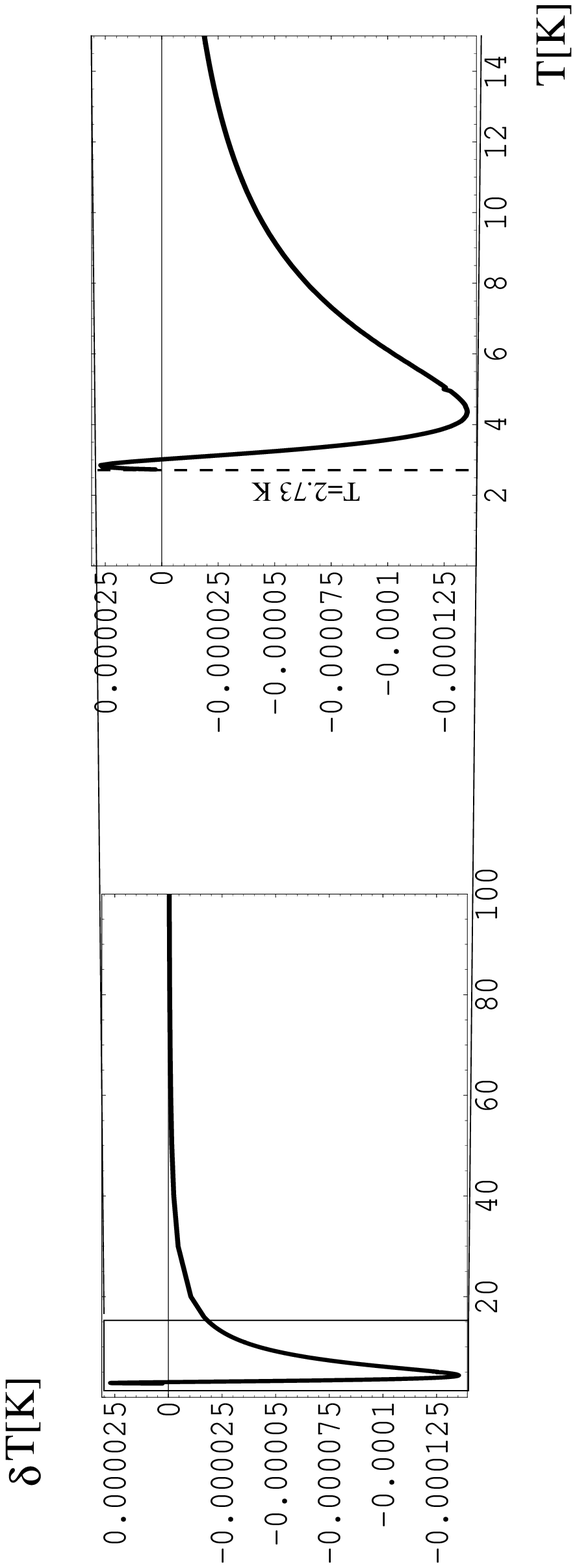}
\end{center}
\caption{Temperature offset $\delta T$ as a function of $T$ 
for $2.73\,\mbox{K}\le\delta T\le 100\,\mbox{K}$ (left panel) and 
$2.73\,\mbox{K}\le\delta T\le 15\,\mbox{K}$ (right panel). The 
offset is defined as $\delta T\equiv\bar{T}-T$ where the temperature 
$\bar{T}$ is extracted by fitting a (dimensionful) 
U(1) BB-spectrum to a (dimensionful) SU(2)$_{\tiny\mbox{CMB}}$ BB-spectrum of temperature $T$ within the 
frequency interval $\nu_{\tiny\mbox{low}}\equiv 1\,\mbox{cm}^{-1}\le\nu\le 10\,T\,\mbox{K$\cdot$cm}^{-1}\equiv 
\nu_{\tiny\mbox{up}}$. For the fit we have used 5000 equidistant frequency points in this interval.}
\label{Fig-3}
\end{figure}

\begin{figure}[tbp]
\begin{center}
\leavevmode
\leavevmode
\vspace{5.3cm} \includegraphics{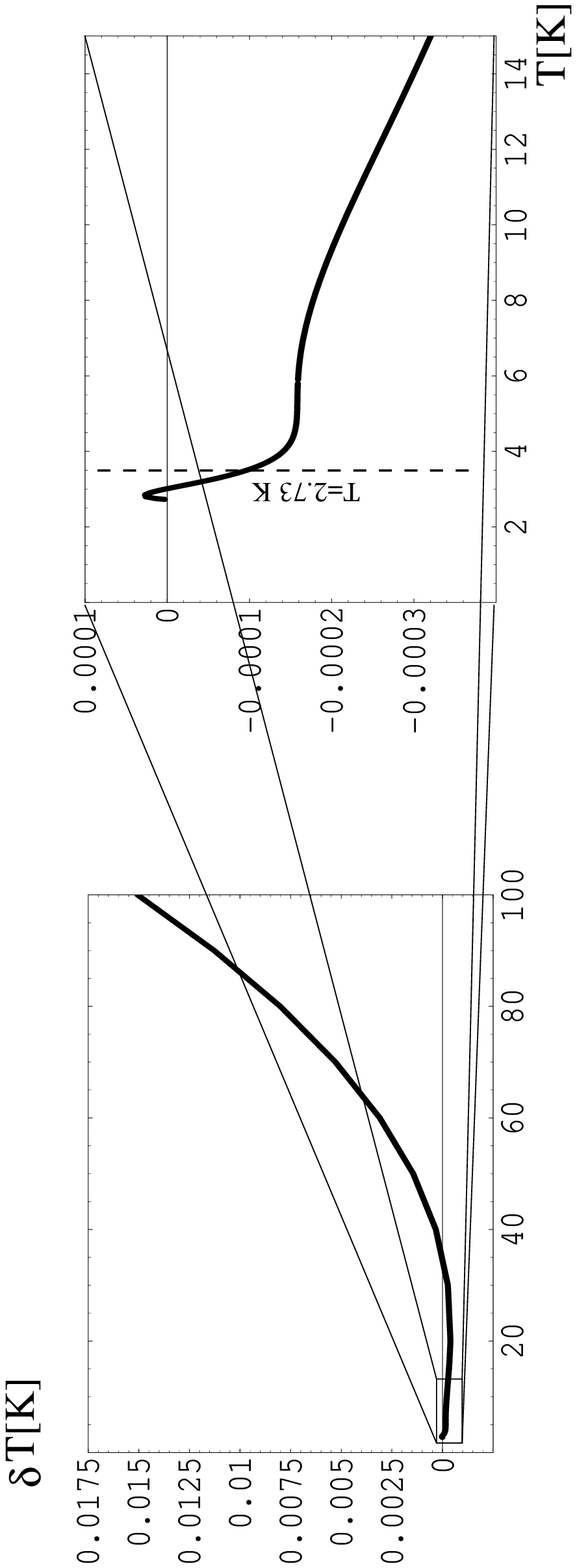}
\end{center}
\caption{Same as in Fig\,\ref{Fig-3} but now keeping $\nu_{\tiny\mbox{up}}$ fixed at 20\,cm$^{-1}$.}  
\label{Fig-3A}
\end{figure}

Let us now discuss the dependence of $\delta T$ on $T$ more quantitatively. Fig.\,\ref{Fig-3} shows the offsets
$\delta T$ as a function of T in the range $2.73\,$K$\le T\le 100\,$K as extracted by fits involving the
{\sl dimensionful} BB spectra\footnote{By dimensionful we mean that the spectral intensities
$I_{\tiny\mbox{U(1)}}(\nu,\bar{T}=T+\delta T)$ and 
$I_{\tiny\mbox{SU(2)$_{\tiny\mbox{CMB}}$}}(\nu,T)$ are divided by $k_B^3$ (expressed 
in SI units). For the fit we have generated 5000 equidistant frequency points covering the 
above-defined range.}. Fig.\,\ref{Fig-3A} shows these offsets when keeping 
$\nu_{\tiny\mbox{up}}=20\,$cm$^{-1}$ fixed as it was done in the 
FIRAS fit. Obviously, there is growing disagreement between the two plots 
for $T>4\,$K. This is explained by the smallness of the frequency 
interval used for generating Fig.\,\ref{Fig-3A}: The fit gives too much weight 
to the BB anomaly at low frequencies. 

Denoting the statistical error of $\delta T$ by $\Delta T$, 
the relative statistical error $\frac{\Delta T}{\delta T}$ of the 
fit grows rapidly with temperature while 
the $\chi^2$/d.o.f. remains comfortably small. 

Let us first discuss the case of Fig.\,\ref{Fig-3}. At $T=5, 7, 10\,$ and 
20\,K we have $\frac{\Delta T}{\delta T}=$25\%, 48\%, 93\%, and 270\% and 
$\chi^2$/d.o.f.=2.6$\times 10^{-4}$, 1$\times 10^{-3}$, 5$\times 10^{-3}$, 
and 4$\times 10^{-2}$, respectively. For $T>80\,$K the $\chi^2$/d.o.f. starts 
to be larger than unity. Judging from the statistical errors, the 
anomaly is only visible up to $T\sim 10$\,K. However, the high quality of the fit, 
which is expressed by $\chi^2$/d.o.f. being substantially smaller 
than unity, persists up to much larger $T$. 

Let us now turn to 
the statistical errors corresponding to the calibration stage 
of FIRAS (Fig.\,\ref{Fig-3A}). Here, we have $\frac{\Delta T}{\delta T}=$25\%, 34\%, 43\%, and 66\% and 
$\chi^2$/d.o.f.=6.6$\times 10^{-4}$, 4.4$\times 10^{-3}$, 2.5$\times 10^{-2}$, 
and 3.9$\times 10^{-1}$ for the same temperatures $T=5, 7, 10\,$, 
and 20\,K, respectively. Thus the fits are well acceptable from the 
statistically point of view. Recall, however, 
that the fitted value for $\delta T$, obtained by keeping
$\nu_{\tiny\mbox{up}}$ fixed at $\nu_{\tiny\mbox{up}}=20\,$cm$^{-1}$, 
is not stable under variations in $\nu_{\tiny\mbox{up}}$. 

\subsubsection{Observation and SU(2)$_{\tiny\mbox{CMB}}$: WMAP} 

Let us now discuss our results adapted to the conditions 
of CMB observations carried out by the WMAP mission. In the Table we give 
the frequency bands used by WMAP \cite{WMAPdoc} for the generation of the  
dipole- and monopole subtracted map of temperature fluctuations over 
the sky.\vspace{0.1cm}\\  
\begin{center}
\begin{tabular}{ccccc}
band & $\nu_{\mathrm{low}}$ [GHz] & $\nu_{\mathrm{up}}$ [GHz] & $\nu_{\mathrm{low}}$ [cm$^{-1}$] & $\nu_{\mathrm{up}}$ [cm$^{-1}$] \\
\hline
K & 19.5 & 25 & 0.65 & 0.83 \\
Ka & 28 & 37 & 0.93 & 1.23 \\
Q & 35 & 46 & 1.17 & 1.53 \\
V & 53 & 69 & 1.77 & 2.30 \\
W & 82 & 106 & 2.74 & 3.54 \\
\end{tabular}
\end{center}
\begin{figure}[tbp]
\begin{center}
\leavevmode
\leavevmode
\vspace{6.5cm} \includegraphics{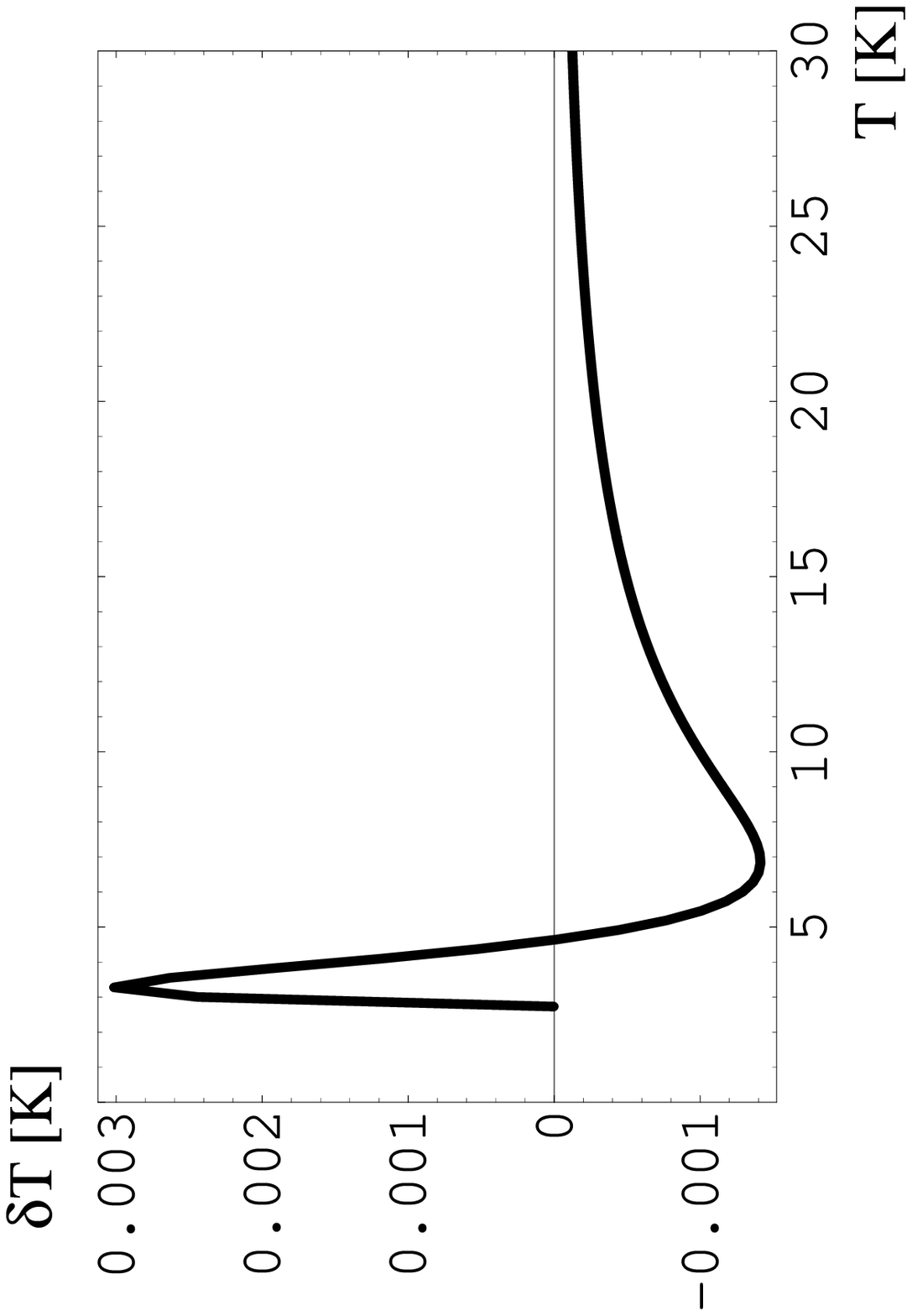}
\end{center}
\caption{Offset $\delta T$ as a function of $T$ 
for $2.73\,\mbox{K}\le\delta T\le 30\,\mbox{K}$. The 
offset is defined as $\delta T\equiv\bar{T}-T$ where the temperature 
$\bar{T}$ is extracted by fitting a (dimensionful) 
U(1) BB-spectrum to a (dimensionful) SU(2)$_{\tiny\mbox{CMB}}$ BB-spectrum of temperature $T$ within the 
frequency of the blueshifted WMAP frequency bands, see text. For the fit we have used 1000 
equidistant points within each of the five WMAP frequency bands.}
\label{Fig-5}
\end{figure}
Our goal is to extract the offsets $\delta T\equiv\bar{T}-T$ in analogy 
to Sec.\,\ref{Firas}. We now perform the fits within all 
blueshifted WMAP frequency bands. Notice that these frequency bands lie deep inside the 
Raleigh-Jeans regime of the BB spectrum. By blueshift we mean that 
the frequencies in the WMAP bands are rescaled 
by a factor of $T/(2.73\,\mbox{K})$. This corresponds to a fictitious 
WMAP observation of the associated BB intensity in an earlier 
Universe at redshift $z=T/(2.73\,\mbox{K})-1$. Within each frequency band we use
1000 equidistant points.

Although a direct comparison of the temperature offsets, extracted under the above conditions, 
with the $TT$ angular correlation functions as extracted from the WMAP temperature 
map is impossible, we may however discuss some qualitative features of
Fig.\,\ref{Fig-5}. First, notice the change in sign at a redshift $z$ of
about $z=1$. This change in sign is accompanied with a strong
gradient. In \cite{SH2007} it was shown that the bulk of the
contribution to the dynamical component to the CMB dipole is 
generated around $z=1$. By inspecting Fig.\,\ref{Fig-5} we observe a
large slope in the $z$-evolution of $\delta T$ around $z=1$. This
implies a rapid built-up of a temperature 
profile out of an initial inhomogeneity at this redshift. Second, notice
the decreasing slope of $\delta T(z)$ for increasing values of 
$z>1$. In this regime the built-up of inhomogeneities is much weaker
than in the vicinity of $z=1$.
\begin{figure}[tbp]
\begin{center}
\leavevmode
\leavevmode
\vspace{6.0cm} \includegraphics{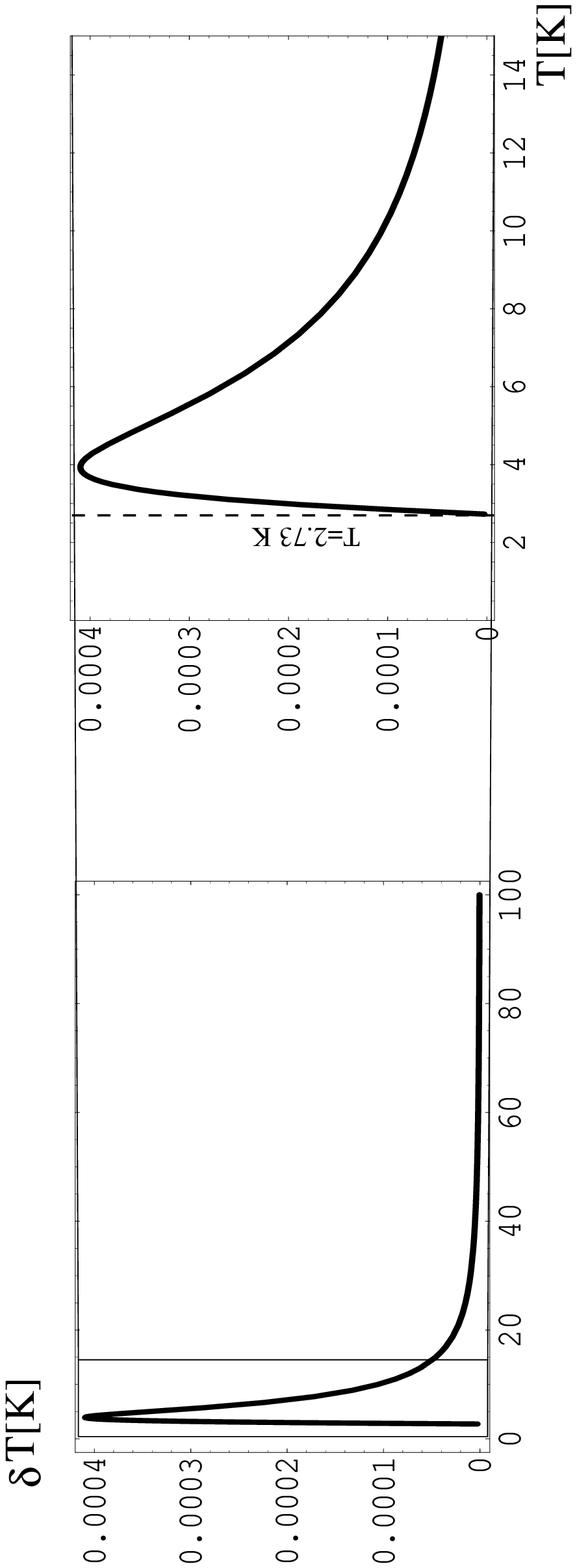}
\end{center}
\caption{Offset $\delta T$ as a function of $T$ 
for $2.73\,\mbox{K}\le\delta T\le 100\,\mbox{K}$ when using the
Stefan-Boltzmann law as a model for the fit of the total energy density. The 
offset is defined as $\delta T\equiv\bar{T}-T$ where $\bar{T}$ is 
the conventional BB temperature attributed to the {\sl integral} over the modified BB spectrum.}
\label{Fig-6}
\end{figure}
Thus there is no major modification of
the primordial spectrum associated with evolution down to $z\sim 1$. Due
to the rapid built-up of the dipole contribution at $z=1$ 
pre-existing, large angle fluctuations are smoothed (inflated) away and get
statistically correlated with the dipole. Based on the work in \cite{SH2007}, we will make this 
assertion much more quantitative in a forthcoming publication \cite{SH2007wp}. 
Finally, notice that the values for $\delta T$ at a given temperature are quite 
different in the case of a FIRAS-like and WMAP-like extraction. For example, at $T=6.83$\,K we have 
$\delta T=-1.44\times 10^{-3}$\,K from WMAP and $\delta T=-0.085\times
10^{-3}$\,K from FIRAS when performing the fit in the range 
$\nu_\mathrm{low}=1\,$cm$^{-1}\le\nu\le 10\,T\,\mbox{K$\cdot$cm}^{-1}\equiv \nu_{\tiny\mbox{up}}$.

\subsubsection{Temperature offsets from Stefan-Boltzmann law\label{STos}}

After having discussed the emergence of temperature offsets $\delta T$ when fitting 
modified to conventional BB spectra we now point out the differences 
of this method as compared to the extraction of $\delta T$ from
integrated spectra (integration over {\sl all} frequencies). The model now is that the Stefan-Boltzmann law 
$\rho=\mbox{const}\,T^4$ shall hold for the integral 
$\rho_{\tiny\mbox{con}}$ over the conventional BB spectrum as well as for the
integral $\rho_{\tiny\mbox{mod}}$ over the modified BB spectrum. 
To linear order in $\delta T$ we thus have
\eqb
\label{modelgl}
\delta T\equiv\bar{T}-T=T\,\frac{\rho_{\tiny\mbox{mod}}-\rho_{\tiny\mbox{con}}}{4\rho_{\tiny\mbox{con}}}\,.
\eqe
In Fig.\,\ref{Fig-6} we show a plot of $\delta T$ as extracted from Eq.\,(\ref{modelgl}). 
Comparing Figs.\,\ref{Fig-6}, \ref{Fig-3} and \ref{Fig-5}, we observe that the sign of 
$\delta T$ does not coincide for sufficiently high temperatures. 
On one hand, the positivity of $\delta T$, as extracted from 
Eq.\,(\ref{modelgl}), is explained by the regime of 
antiscreening having a larger weight in the integration of the modified BB spectrum than the 
regime of screening. On the other hand, the fit to the BB {\sl spectrum} interpretes the regime of 
antiscreening in terms of a {\sl decrease} of temperature in accord with Wien's displacement law. 
So the two methods produce even qualitatively different results.

\section{Summary and Conclusions\label{SC}}

In this article we have performed an analysis of offsets between radiation- and wall-temperatures
in ideal black bodies. On the theoretical side these offsets arise due to nonabelian effects of an SU(2) Yang-Mills
theory of scale $\Lambda=10^{-4}$\,eV
\cite{SHG2006,SHG2006BB,RH2005B,GH2005}: SU(2)$_{\tiny\mbox{CMB}}$. The
deviation from the conventional black-body spectra 
peak at temperatures a few times that of the present cosmic microwave background. There are two ways of extracting these temperature 
offsets: Either assume a Stefan-Boltzmann form of the energy density
(integrated, modified black-body spectrum) 
or assume a conventional black-body spectral shape as fit models for the
data representing 
the modified spectrum. For the frequency ranges and the frequency bands used in the FIRAS instrument calibration and in the WMAP mission,
respectively, we have extracted these offsets using the latter method. We find that in 
the case of FIRAS the nonabelian effects, as predicted by SU(2)$_{\tiny\mbox{CMB}}$, 
are too small to rise above the 
errors of the experiment. We give qualitative explanations for the global, 
negative offset of about $-4\,$mK 
observed for in-flight calibration at temperatures larger than 
$T_{\tiny\mbox{CMB}}=2.73\,$K  and the positive offset for 
calibration at $T=2.2\,$K. Both effects are further evidence toward the correctness of the postulate
that photon {\sl propagation} is described SU(2)$_{\tiny\mbox{CMB}}$. In order to verify or falsify
this postulate a terrestrial low-temperature, low-frequency, high-precision measurement of the black body spectral
intensity surely is required. We are confident, however, that work
performed in the framework of \cite{SH2007} will yield the 
suppression and correlation of the low-lying CMB multipoles
\cite{SH2007wp} as extractable from the data of the WMAP mission \cite{Schwarz2007}.

Comparing our result for the FIRAS and the WMAP setting, we obtain
offsets that are larger by an order of magnitude for the spectral fits in
the blueshifted frequency bands of the latter. 
Interestingly, the value of the minimum in the WMAP fit, which occurs at a temperature 
corresponding to a redshift $z=1.5$, is close in magnitude to the amplitude of the (dynamical component of the) 
dipole, for a discussion see \cite{SH2007}. Furthermore, the large slope
of $\delta T(z)$ in the vicinity of $z=1$ very likely is associated
with the rapid built-up of the dynamical component of the CMB
dipole. This process implies a smoothing of primordial, large-angle
fluctuations and a correlation thereof. We will make this assertion much
more quantitative in a forthcoming publication \cite{SH2007wp}.  

Finally, for the extraction of $\delta T$ from presuming the Stefan-Boltzmann law we
observe vast deviations compared to the extraction from spectral fits. This is
true for both magnitude and sign. Thus to speak of temperature offsets
generated by whatever possible mechanism it is imperative to 
specify the method used to extract them.

\section*{Acknowledgments}

We would like to thank Frans Klinkhamer for useful discussions. One of
us (R.H.) would like to thank Richard Battye for a stimulating pub
conversation during a cosmology conference held at Imperial College in
March. Many thanks to the organizers of this conference also for their financial
support.

\end{document}